\documentclass{ws-ijmpa}
\usepackage[super,compress]{cite}
\usepackage{graphicx}
\usepackage{xcolor}
\usepackage{url}
\usepackage{enumerate} 
\usepackage{xcolor}

\usepackage{graphicx}
\usepackage{dcolumn}
\usepackage{bm}
\usepackage[super]{natbib}
\begin{document}
\markboth{Zuzanna Chochulska}{Prospects of angular correlation studies of identified hadrons}

%
\catchline{}{}{}{}{}
%

\title{Prospects of angular correlation studies of identified hadrons in the LHC Run 3 with the ALICE experiment
}

\author{Zuzanna Anna Chochulska
}

\address{Faculty of Physics, Warsaw University of Technology, Koszykowa 75\\ 
Warsaw, 00-662, Poland\\
zuzanna.chochulska.dokt@pw.edu.pl}

\maketitle


\begin{abstract}
\noindent The study of angular correlations of identified hadrons makes it possible to understand the impact of different processes that contribute to the hadronization mechanism in heavy-ion collisions and small collision systems. Depending on the quark composition and the system properties, the angular correlations exhibit different shapes.
\newline
The mechanism of hadron production is non-perturbative, and because of that, only phenomenological models can be used. Those models still fail to fully describe hadron production. In particular, angular correlations for like-sign baryons are not reproduced by the models. The upgraded ALICE detector and the new analysis framework allow for precise measurements of baryon angular correlations which provide stronger constraints on the hadronization mechanisms. 

\keywords{CERN; ALICE; O2 software.}
\end{abstract}

\ccode{PACS numbers:}


\section{Introduction}
Hadronization is a process of forming hadrons from quarks and gluons. It is a non-perturbative mechanism that can only be described using phenomenological models. The first models describing these phenomena date back to the 1970s and are still being developed and improved. 

\begin{figure} 
        \centering
        \includegraphics[width=1.0 \linewidth]{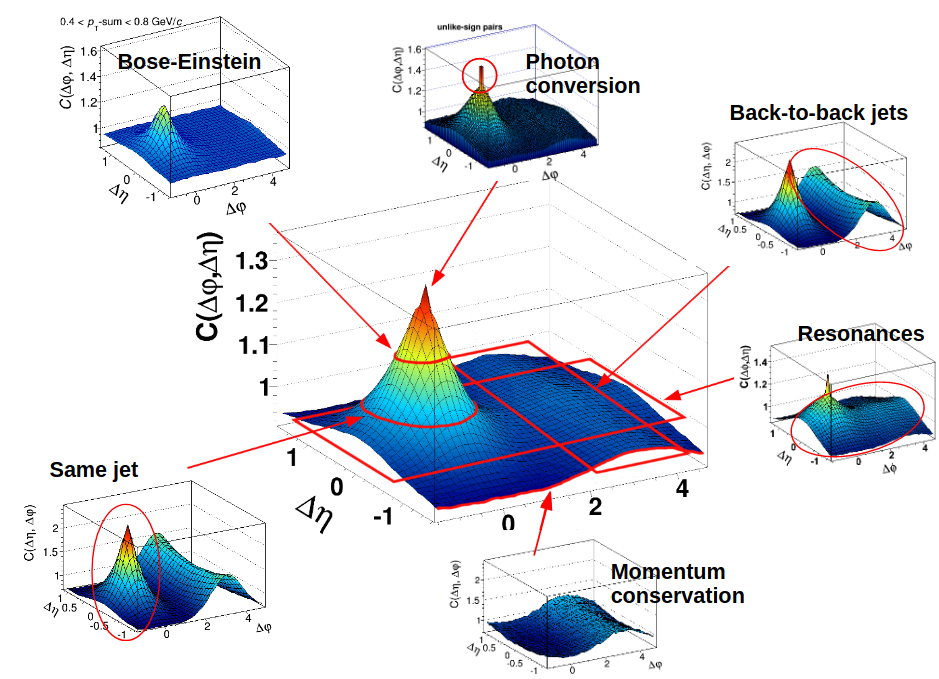}
        \caption{Components of an angular correlation function \cite{Obrazek_correlation}.}
        \label{fig: Ang Corr Factors}
    \end{figure}

A powerful tool used to describe this process are angular correlation functions. The effects that comprise the properties studied include resonance decays, energy and momentum conservation, and Coulomb interactions. The components that constitute the typical angular correlation function are shown in Fig. \ref{fig: Ang Corr Factors}.
    
The angular correlation function is defined as:

\begin{equation}
    C (\Delta \eta, \Delta \varphi) = \frac{N^{\mathrm{mixed\ pairs}}}{N^{\mathrm{signal\ pairs}}}\frac{S(\Delta \eta, \Delta \varphi)}{B(\Delta \eta, \Delta \varphi)},
    \label{eq correlation function probability}
\end{equation}
where: 
\begin{itemize}
    \item $S(\Delta \eta, \Delta \varphi)$ is the signal distribution, 
    \item  $B(\Delta \eta, \Delta \varphi)$ is the background distribution.
\end{itemize}

\noindent To construct the signal and background distributions one needs to correlate particles as a function of their difference in the pseudorapidity $\Delta \eta$ and in the azimuthal angle $\Delta \varphi$. Pairs coming from the same event make up the signal distribution and pairs coming from different events contribute to the background distribution. Events used for constructing the background distribution are chosen via an event-mixing algorithm. The background distribution enables us to take into account the detector acceptance effects.
\newline 

\noindent The first angular correlation functions were studied in the 1980s, when an anticorrelation in the vicinity of the (0,0) point of the distribution was first reported for the same baryon number pairs. Back then, the possible explanation for this phenomenon was the energy limitation. This was believed to be causing the anticorrelation due to the fact that protons had a considerable mass for the collision energy available, which was just $\sqrt{s} = 29\, \mathrm{GeV}$. However, this limitation quickly ceased to play a key role.

Measurements of angular correlations for identified hadrons were performed by the ALICE Collaboration in proton--proton collisions at $\sqrt{s} = 7\, \mathrm{TeV}$ and the same shape persisted~\cite{Insight_correlations_MC_data}. 

\begin{figure}
    \centering
    \includegraphics[width = 0.9\textwidth]{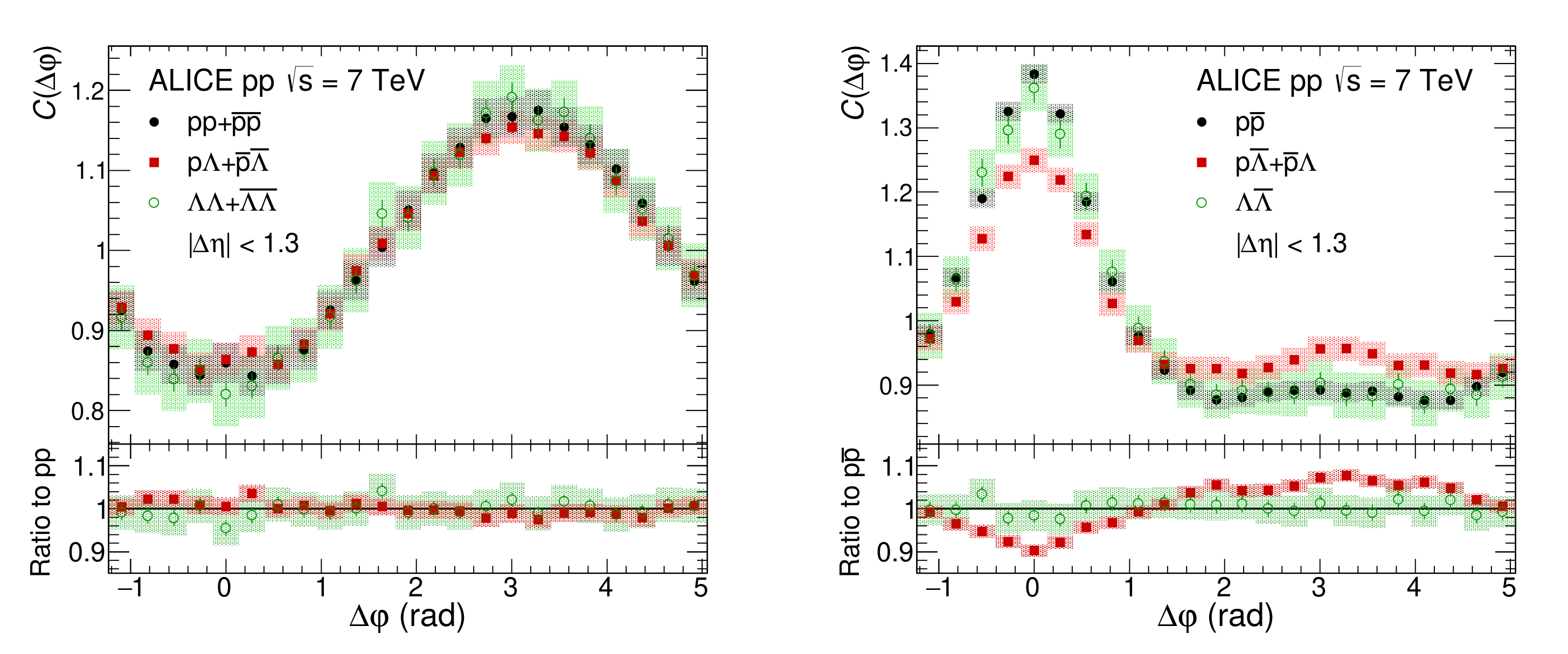}
    \caption{$\Delta \eta$ integrated projections of correlation functions for combined pairs of (left) pp + $\Bar{p}\Bar{p}$, p$\Lambda$ + $\Bar{p}\Bar{\Lambda}$, and $\Lambda$$\Lambda$ + $\Bar{\Lambda}$$\Bar{\Lambda}$ and (right) p$\Bar{p}$, p$\Bar{\Lambda}$ + $\Bar{p}\Lambda$, and $\Lambda$$\Bar{\Lambda}$. Statistical (systematic) uncertainties are represented by bars (boxes). Figure taken from Ref.~\cite{Insight_correlations_MC_data}.}
    \label{fig:Insight_correlations_MC_data}
\end{figure}

Figure \ref{fig:Insight_correlations_MC_data} shows projections of angular correlations of selected baryon pairs on the $\Delta \varphi$ axis. On the left, one can observe an anticorrelation for the like-sign baryons that cannot be explained by any of the Monte Carlo models used so far, namely PYTHIA6 Perugia-0, PYTHIA6 Perugia-2011, PYTHIA8
Monash, and PHOJET~\cite{Insight_correlations_MC_data}. On the other hand, a correlation is observed for the unlike-sign pairs, which can be seen in the right panel of Fig. \ref{fig:Insight_correlations_MC_data}. The models agree with the results of unlike-sign pairs, but the observed peak near the (0, 0) is overestimated.
The STAR Collaboration performed a similar angular correlation analysis \cite{PhysRevC.101.014916} and also obtained an anticorrelation around the (0, 0) point for the like-sign baryons. Since then, many hypotheses of why one can observe anticorrelation have been posed. Does particle mass matter? Does their flavour matter? How does the observed anticorrelation change with the multiplicity of the collision? 

These and  other questions should find answers with the new data collected by the ALICE detector in Runs 3 and 4. The upgraded ALICE detector with its acquisition rate increased by up to a factor of about 50 in Pb--Pb collisions may allow new analyses to be performed due to the large data samples.

\section{ALICE}
\textbf{ALICE} (\textbf{A} \textbf{L}arge \textbf{I}on \textbf{C}ollider \textbf{E}xperiment) \cite{ALICE_experiment} is one of the four main experiments at the \textbf{L}arge \textbf{H}adron \textbf{C}ollider (LHC)~\cite{LHC_machine}. After Long Shutdown 2 (2018--2022) the rate of collisions increased and because of that data-taking mode has been switched to continuous, with the unit of data lasting 10 ms being called TimeFrame (\textbf{TF}). Furthermore, several subdetectors have been majorly upgraded. New gas electron multipliers replaced old wire chambers to increase the readout rate in the Time Projection Chamber (\textbf{TPC}), a new Inner Tracking System (\textbf{ITS}) and new Muon Forward Tracker (MFT) have been installed. A completely new Fast Interaction Trigger (\textbf{FIT}) is used to trigger other detectors, and it enables measurements of luminosity and forward multiplicity, and provides fast timing used for Time-of-Flight (\textbf{TOF}).

Apart from significant hardware upgrade, ALICE has also upgraded its software and the data format used. The new data files are constructed from flat Apache Arrow~\cite{Appache_Arrow_webpage} tables, where each row represents a different entity such as a collision or a track. Appache Arrow is a language-agnostic and open-source columnar in-memory data representation format, that is designed for efficient and high-performance analytics.

Furthermore, the new ALICE $O^2$ (Online-Offline) software \cite{Buncic:2015ari} enables reconstruction, calibration, and simulation of the ALICE data. The $O^2$ uses declarative programming which differs from the imperative one in the sense that it expresses the logic of computation by describing what the program should achieve and not how it should do it. 

\subsection{FemtoUniverse}
FemtoUniverse is a package in $O^2$ developed by the Warsaw University of Technology group and is dedicated to femtoscopic analyses and angular correlation analyses of identified hadrons. 

\begin{figure}
    \centering
    \includegraphics[width=0.7\textwidth]{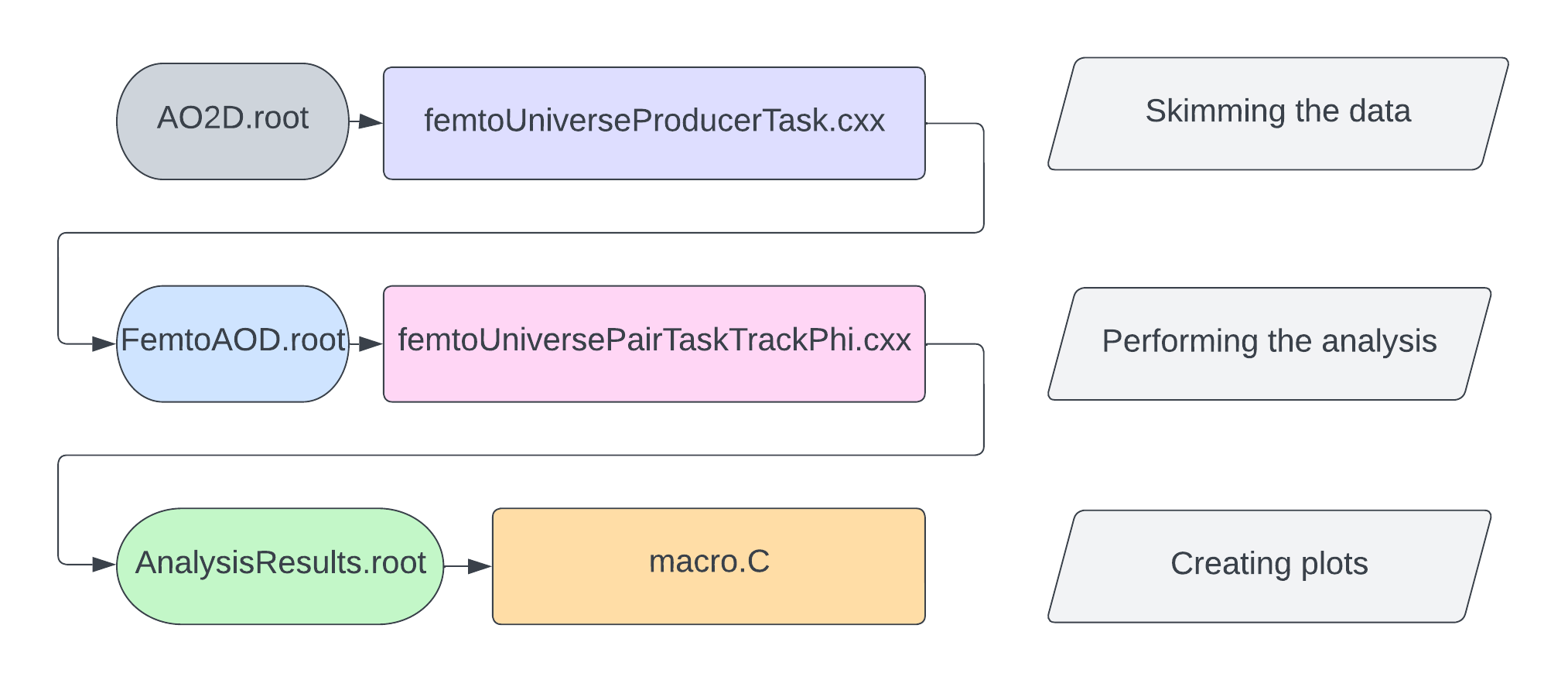}
    \caption{The simplified code flow for most FemtoUniverse analyses.}
    \label{fig:FemtoUniverse_flow}
\end{figure}

What makes this package special is that it can be divided into two phases: generation of the derived data and running the analysis task. The simplified code flow for the typical FemtoUniverse analysis is presented in Fig. \ref{fig:FemtoUniverse_flow}.
The first step enables to reduce the data of all unnecessary parts (unwanted variables, rejected collisions, and particles), but also allows to reconstruct quickly decaying particles, which are impossible to register using detectors, such as $\phi$ mesons, V0s and D mesons.

\section{Tools}
\subsection{GRID}
To meet the high demand on data storage and data processing the Worldwide LHC Computing Grid (\textbf{WLCG})~\cite{GRID_project} has been created. It is a distributed network of connected computational sites, which provides means to store, process, and analyze the  LHC data. The GRID establishes connections between distributed computing resources, data services, and processor farms over the Internet, providing standard protocols for program execution and output download.

\subsection{Hyperloop}
Analyzers can use so-called train systems, which allow the use of up to a few hundred TB of data.
The Hyperloop train system~\cite{Quishpe:2021pls} has been developed to handle the large data samples from Run 3 and 4 using the \textbf{GRID} infrastructure. Hyperloop is presented in the form of a website, which GUI is based on the React.js library. It uses a PostgreSQL database for storage and a Java-based model for bookkeeping of wagons and datasets.

\section{Outlook}

The FemtoUniverse package, developed by the group from Warsaw, allows analyzers to obtain both angular correlation functions and femtoscopic correlations for identified hadrons. At this point, most of the analyses being conducted are in the initial stages, however, further studies are planned, for which additional development of the $O^2$ software will be necessary. The new software used in the ALICE experiment presents itself with a full range of capabilities to perform analyses on large data samples using the tools described above.
This research will allow the study of angular correlations that will provide a deeper understanding of the hadronization process. In particular, conducting an analysis of angular p$\phi$ correlations will allow us to see if the resulting anticorrelation is affected by the mass of the particle.


\section{Acknowledgements}
I would like to thank the ALICE Collaboration, for access to the state-of-the-art analysis software and data, as well as for the tremendous support I have received over the years.

This work was supported by the Polish National Science Centre
under the agreement no. UMO-2021/43/D/ST2/02214 and The Polish Ministry of Science and Higher Education under agreements no. 2022/WK/01, 2023/WK/07, and 5236/CERN/2022/0.


\end{document}